\documentclass[a4paper,twocolumn,floatfix]{revtex4}
\usepackage{graphicx}

\begin{document}

\title{Correlation between molecular orbitals and doping dependence of the electrical conductivity in electron-doped Metal-Phthalocyanine
compounds}

\author{M. F. Craciun}
\author{S. Rogge}
\author{A. F. Morpurgo}

\affiliation{Kavli Institute of Nanoscience, Delft University of
Technology, Lorentzweg~1, 2628\,CJ Delft, The Netherlands}

\begin{abstract}
We have performed a comparative study of the electronic properties
of six different electron-doped metal phthalocyanine (MPc)
compounds (ZnPc, CuPc, NiPc, CoPc, FePc, and MnPc), in which the
electron density is controlled by means of potassium
intercalation. In spite of the complexity of these systems, we
find that the nature of the underlying molecular orbitals produce
observable effects in the doping dependence of the electrical
conductivity of the materials. For all the MPc's in which the
added electrons are expected to occupy orbitals centered on the
ligands (ZnPc, CuPc, and NiPc), the doping dependence of the
conductivity has an essentially identical shape. This shape is
different from that observed in MPc materials in which electrons
are also added to orbitals centered on the metal atom (CoPc, FePc,
and MnPc). The observed relation between the macroscopic
electronic properties of the MPc compounds and the properties of
the molecular orbitals of the constituent molecules, clearly
indicates the richness of the alkali-doped metal-phthalocyanines
as a model class of compounds for the investigation of the
electronic properties of molecular systems.

\end{abstract}

\maketitle

The electronic properties of organic molecular solids are
characterized by narrow electronic bands originating from the
weak, non-covalent intramolecular bonds that hold the materials
together. The narrowness of the bands results in an inter-band
separation that is typically much larger than the bandwidth, so
that the character of each band is closely related to that of the
molecular orbital from which the band originates. As the band
structure plays an important role in determining the electronic
properties of a material, the nature of the underlying molecular
orbitals can be expected to produce observable effects in the
electrical conductivity of the materials. However, it is unclear
whether other phenomena (e.g., electron-electron and
electron-phonon interactions or structural effects) that also play
an important role in determining the electronic properties of
materials can mask the effects originating from the details of the
molecular orbitals.

Here we address this issue through a comparative study of the
doping dependence of the electrical conductivity of six different
metal phthalocyanine (MPc) compounds (ZnPc, CuPc, NiPc, CoPc,
FePc, and MnPc), in which the electron density is controlled by
means of potassium intercalation. We find that for all MPc's in
which the added electrons transferred from the Potassium atoms are
expected to occupy orbitals centered on the ligands (ZnPc, CuPc,
and NiPc), the doping dependence of the conductivity has an
essentially identical shape. This shape is different from that
observed in MPc materials in which electrons are also added to
orbitals centered on the metal atom (CoPc, FePc, and MnPc). We
conclude that in MPc compounds the characteristics of the
molecular orbitals of individual molecules are directly visible in
the electrical conductivity of the materials.

\begin{figure}[ht]
  \centering
  \includegraphics[width=0.7\columnwidth]{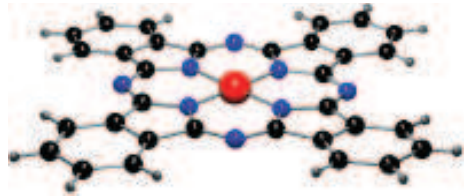}
\caption{Molecular structure of metal phthalocyanines: they
consist of a ring of Carbon, Nitrogen and Hydrogen atoms,
surrounding a metallic ion. }
  \label{fig:MPc}
\end{figure}

MPc's form a large class of organic molecules \cite{MPcbook} that
are ideally suited to investigate the relation between electrical
conductivity and molecular orbitals. They consist of a metal atom
located at the center of a planar ligand shell formed by Carbon,
Nitrogen and Hydrogen atoms (see Fig. \ref{fig:MPc}), with the
metal atom determining the energy and the degeneracy of the
molecular orbitals \cite{orbitals,STM1,STM2}. The different
individual MPc molecules are nearly isostructural, and also their
crystal structure only exhibits minor differences: this makes
structural effects an unlikely origin of differences in the
conductivity of different MPc compounds. Finally, MPc's have been
subject of thorough investigations in the past
\cite{MPcreview1,MPcreview2} and much is known about their
electronic properties, which facilitate the rationalization of the
experimental observations.

Doping MPc materials with electrons rather than with holes is
crucial for our investigations for two main reasons. Firstly, in
hole-doped MPc compounds past systematic investigations have shown
that the holes always reside on the same molecular orbital
centered on the ligands. This prevents the possibility to vary the
molecular orbital occupied by the charge carriers responsible for
electrical conductivity. On the contrary, in electron-doped MPc's
the electron can either occupy a ligand or a metal orbital
depending on the specific molecule considered \cite{orbitals}.
Secondly, it is known that it is possible to transfer a larger
amount of charge by reducing \cite{Clack,Taube} rather than by
oxidizing MPc's, so that electron doping gives experimental access
to a much larger interval of charge density as compared to hole
doping \cite{MPcreview1,MPcreview2}. So far, however,
electron-doped molecular compounds have remained vastly
unexplored, probably because their sensitivity to oxidizing agents
increases the technical difficulties involved in their
investigation.

Our work is based on thin-film (20nm thick) materials that are
thermally evaporated on the surface of a silicon-on-insulator
(SOI) substrate. Charge carriers are introduced in these films by
means of chemical doping with Potassium atoms. All the steps of
our investigations, including film deposition, doping, and
electrical transport measurements, have been carried out at room
temperature, {\it in-situ} in a single ultra-high vacuum (UHV)
system with a base pressure $< 5 \ 10^{-11}$\,mbar. This prevents
the occurrence of visible degradation of the doped films over a
period of days.

\begin{figure}[ht]
  \centering
  \includegraphics[width=1\columnwidth]{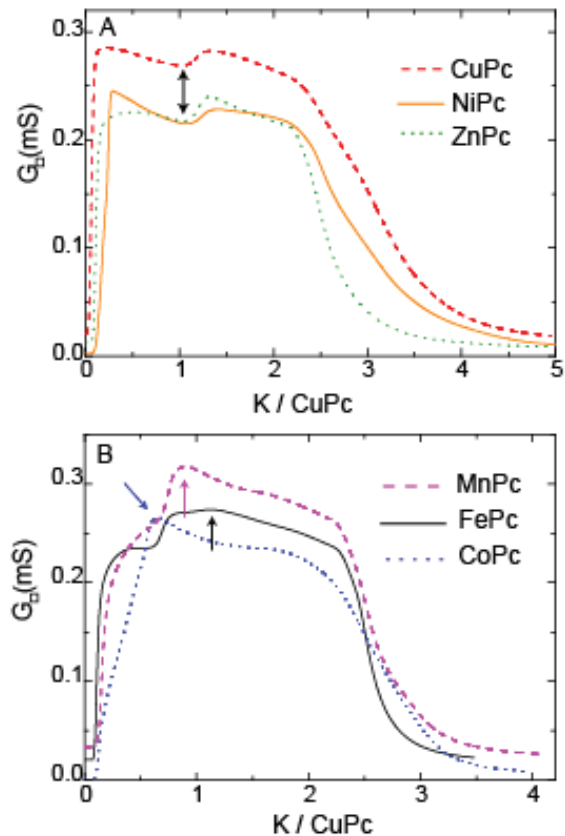}
\caption{Square conductance measured at room temperature as a
function of potassium concentration, on 20 nm thick films
deposited on Si (001) of CuPc, NiPc and ZnPc (A), and of CoPc,
FePc, and MnPc (B). }
  \label{fig:doping-curve}
\end{figure}

Figure \ref{fig:doping-curve}A and B show the conductance of the
films of six different MPc's as a function of doping
concentration. The data are obtained by measuring the conductance
while exposing the film to a constant flux of K atoms generated by
a current-heated getter source. In order to determine the
potassium concentration in the film, we have performed an
elemental analysis for several doping levels using {\it ex-situ}
RBS for CuPc. We have then used the K-CuPc data to scale the
concentration of the other molecular films as a function of
potassium exposure time. Although the absolute determination of
the potassium concentration is affected by a relatively large
uncertainty (approximately 1 K atom per molecule at high doping
density), these measurements indicate that electrons transferred
from the potassium atoms to the molecules are enough to fill one
or more molecular orbitals, depending on the orbital degeneracy.
This is consistent with past studies \cite{orbitals} that have
shown the possibility to add at least four electrons to most of
the MPc's used in our work.

We first summarize the similarities in the behavior of the
conductivity of the different K-MPc compounds that have been
discussed in our recent work \cite {AdvMat}. For all the materials
the conductance first increases with potassium concentration up to
a high value that is comparable for the different molecules, it
remains high in a broad range of concentrations, and it eventually
decreases to the level observed for the pristine material. The
temperature dependence of the conductance shows that all the
materials are metallic in the highly conducting state and
insulating in the low conductivity regions at low and very high
doping. We have also shown that well-defined intercalated phases
exist and that Raman studies confirm the occurrence of electron
transfer from the Potassium atoms to the molecules.

We focus on the similarities and differences between the shape of
the doping curve, i.e. the doping dependence of the conductivity,
measured for the different MPc compounds, that have not been
previously discussed. Measurements on more than 200 films
demonstrate that these differences are experimentally robust. Fig.
\ref{fig:doping-curve}A shows that for CuPc, NiPc and ZnPc the
behavior of the electrical conductance upon doping is very
similar. In particular, the high conductivity part of the doping
curve consists of two regions, which extend over the same
concentration range. These two regions are separated by a shallow
minimum, which occurs at the same doping concentration for all
these three systems (indicated by the arrows in Fig.
\ref{fig:doping-curve}A). This profile of the doping curve is
different from that observed for films of CoPc, FePc and MnPc (see
Fig. \ref{fig:doping-curve}B). For these systems, the doping curve
exhibits only one conductance maximum, which is located at
different concentrations for the three different molecules, as
shown by the arrows in Fig. \ref{fig:doping-curve}bB. In addition
the precise structure of the doping curve for these three
molecules is different. Specifically FePc and MnPc have a similar
initial increase in the conductance upon doping, whereas the
initial conductance increase is slower for CoPc. Further the
conductance of MnPc continue to increase and changes the slope
before reaching the maximum, whereas the increase in conductance
of the FePc exhibits two steps before reaching the highest
point.\\

The similarities and differences in the doping curves correlate
well with the known way in which electrons fill the molecular
orbitals of individual MPc's molecules, upon reduction. Recent
theoretical calculations \cite{orbitals} of the electronic
structure of Metal Phthalocyanines, in agreement with
spectroscopic observations \cite{Clack}, indicate that for all the
four subsequent reduction steps of CuPc, NiPc and ZnPc the
electrons fill the same, doubly degenerate 2e$_{g}$ orbital,
belonging to the ligand shell. This identical orbital filling
rationalizes the identical shape of the doping curve that is
observed experimentally for these three molecules upon increasing
the density of charge carriers. On the contrary for CoPc and FePc,
the calculations show that, respectively, the first one and two
electrons fill orbitals centered on the central metal atom. Only
after these orbitals have been filled, electrons are added to the
2e$_{g}$ orbital of the ligand. This is why for CoPc and FePc
films the shape of the doping curves are different from those of
CuPc, NiPc, and ZnPc. It is also why the doping curves of CoPc and
FePc are different among themselves, since for the first molecule
only one electron is added to the 1e$_{g}$ orbital, with strong d
character \cite{orbitals} whereas for the second molecule, two
electrons fill the 1e$_{g}$ and a$_{1g}$ metal orbitals. For MnPc,
reliable ab-initio calculations are not available. However, based
on the analysis of Ref. 2, we expect that the reduction steps with
which the first three electrons are added to the molecule should
involve orbitals with metal character. Thus, for MnPc, we expect a
doping curve different from that of all other molecules and
possibly exhibiting some similarity to that of FePc, in which two
electrons are added to orbitals centered on the metal. The
comparison of the MnPc and FePc doping curves in Fig.
\ref{fig:doping-curve}B show that this is indeed the case.\\

In summary, we have observed that the doping dependence of
electrical conductivity for six different electron-doped MPc
compounds correlates with the orbitals involved in the reduction
of the individual molecules. This provides a direct experimental
demonstration of the role of molecular orbitals in determining the
macroscopic properties of molecular materials. It is remarkable
that the effect is clearly visible experimentally, in spite of the
many other physical process and phenomena that determine the
conductance of molecular systems in the solid state, such as for
instance electron-electron and electron-phonon interaction, doping
inhomogeneity, etc. The observed relation between doping-dependent
conductivity and orbitals of the constituent molecules indicate
that electron-doped metal phthalocyanine compounds are a unique
class of materials for the systematic study of the electronic
properties of molecular systems.\\

We are grateful to Y. Iwasa, S. Margadonna, and K. Prassides for
useful discussions and collaboration. Financial support from FOM,
KNAW, and the NWO Vernieuwingsimpuls 2000 program is acknowledged.

\end{document}